\documentclass[reprint, showpacs, showkeys, nofootinbib, aps, prd]{revtex4-1}
\usepackage{bm}
\usepackage{amsmath}
\usepackage{mathptmx}
\usepackage{amsfonts}
\usepackage{amssymb}
\usepackage[pdftex]{graphicx}
\usepackage{color}
\usepackage{mathptmx}
\RequirePackage[colorlinks,citecolor=darkred,urlcolor=magenta,linkcolor=darkblue,bookmarks=false]{hyperref}
\usepackage{subfig}
\usepackage[toc,title]{appendix}
\usepackage[misc]{ifsym}
\usepackage{adjustbox}
\usepackage{multirow}
\usepackage{makecell}

\allowdisplaybreaks

\definecolor{darkblue}{rgb}{0.0, 0.0, 0.62}
\definecolor{deepmagenta}{rgb}{0.8, 0.0, 0.7}
\definecolor{darkred}{rgb}{0.55, 0.0, 0.0}

\RequirePackage{mymacro}

\RequirePackage{mymacro}
   \graphicspath{{./graphs/}}

\begin{document}
\title{Perturbation in an interacting dark Universe}
\author{Srijita Sinha}
\email{sinha.srijita@niser.ac.in}
\affiliation{School of Physical Sciences, National Institute of Science Education and Research, Jatni 752050, Odisha, India}
\affiliation{Indian Institute of Science Education and Research Kolkata, Mohanpur, Nadia, 741246, India}
\author{Manisha Banerjee}
\email{banerjee.manisha717@gmail.com} 
\affiliation{Department of Physics, Visva-Bharati, Santiniketan -731235, India}
\author{Sudipta Das}
\email{sudipta.das@visva-bharati.ac.in}
\affiliation{Department of Physics, Visva-Bharati, Santiniketan -731235, India}
\begin{abstract}
In  this  paper  we  have  considered  an  interacting  model of dark energy and have looked into the evolution of the dark sectors. By solving the perturbation equations numerically, we have studied the imprints on the growth of matter as well as dark energy fluctuations. It has been found that for higher rate of interaction strength for the coupling term, visible imprints on the dark energy density fluctuations are observed at the early epochs of evolution.
\end{abstract}


\maketitle
\section{Introduction} \label{sec:intro}
The high precision cosmological observations \cite{SupernovaSearchTeam:1998fmf, SupernovaCosmologyProject:1998vns,  Meszaros:2002np, Planck:2014loa, ahn2012ninth} strongly suggest that the Universe is expanding, and surprisingly the nature of expansion is accelerating. Data also reveals that the onset of this acceleration is a recent phenomenon and has started at around $z \sim 0.5$~\cite{reiss2001zt, LAmendola2003zt}. Considering Einstein's theory of general relativity to be the correct theory describing our Universe, the driving agent behind the accelerated expansion, referred to as ``dark energy'' (DE),  should have sufficient negative pressure in order to counter balance the force of gravity \cite{padmanabhan2006dark}. Undoubtedly, the cosmological constant $\Lambda$ appears to be the simplest and the most successful of all candidates, but is troubled by the discrepancy arising from the mismatch in theoretical prediction and observational requirement. An alternative to cosmological constant model are dynamical DE models, preferably driven by a scalar field with a potential.  There has been a profusion of proposals for DE models, such as quintessence, k-essence, phantom, chaplygin gas, tachyon models, holographic DE models and so on (see  \cite{sahni2006reconstructing, bamba2012dark,armendariz2001essentials, caldwell2002phantom, carroll2003can, kamenshchik2001alternative, sen2002tachyon, padmanabhan2002accelerated, copeland2006dynamics, amendola2010dark, sinha2021jcap} and the references therein). Despite having their own merits and demerits, most of these models happen to be more or less consistent with current observations. Hence the origin and true nature of DE remain a mystery, and the search is still on for a viable model of DE. Along this line, recently, a scalar field model of DE has been proposed by Das \etal \cite{Das2018RAA}, where it has been shown that a simple, functional dependence of the energy density of the dark energy sector leads to a double exponential potential and can have several interesting cosmological implications. The double exponential potential models are very well-studied in the context of inflation as well as dark energy \cite{Barreiroprd2000, RubanoGRG2002, SenPLB2002}. They have been ornamental in solving a number of cosmological problems related to early time and late-time cosmology. 
 
Usually the dark energy component is considered to be non-interacting and  is assumed to be non-clustering because of its anti-gravitating properties. However, as the nature of dark energy is still unknown, an interaction between various constituent components of the Universe cannot be ruled out and may provide a more general scenario. These interacting models can be helpful in alleviating the cosmological coincidence problem which seeks the reason for the comparable energy densities of the  dark matter (DM) and dark energy sectors at the present epoch despite of having completely different evolution histories. In fact, an interaction between the dark energy and dark matter components have been studied in a number of dark energy models and found to be useful in solving the coincidence problem  \cite{zimdahl2001PLB, AP2000PRD, Yuri2015IJMPD, yang2018prd,mukherjee2017search, pavon2004jcap,SD2014ASS,sinha2020epjp,sinha2021prd}. 

Keeping in mind the above facts, we consider a modification of the scalar field model of DE described in \cite{Das2018RAA} and introduce a coupling term through which the dark sectors are allowed to interact among themselves. As there is no theoretically preferred form of the interaction term, we make a simple and popular choice for the form of interaction and look into the perturbative aspects of this phenomenological choice. The presence of a DE component is expected to slow down the rate of structure formation of the Universe because of its repulsive gravity effect and hence should have its imprint on the growth of perturbations. The upshot of different dynamical DE models on the structure formation of the Universe will be different as they evolve differently \cite{salvatelli2014prl, banerjee2021growth, yang2018jcap, yang2018prd}. The nonzero interaction term will further affect the evolution of the dark sectors and hence should leave visible imprints on the growth of perturbations depending on the strength of the coupling term. In \cite{Das2018RAA}, the authors have studied the perturbative effect of this particular DE model where the dark sectors were allowed to evolve independently. We are interested to know how the growth of structure gets affected for the DE model described in \cite{Das2018RAA} in the presence of a dynamical coupling term. In this work, the perturbation equations have been solved using the publicly available Boltzmann code \camb, and the effect on the growth of matter and dark energy components has been studied. We have tested the interacting model with different observational datasets like the cosmic microwave background (CMB)~\cite{planck2018cp}, baryon acoustic oscillation (BAO)~\cite{beutler2011mnras, ross2015mnras, alam2017sdss3}, Type Ia Supernovae (SNe Ia)~\cite{scolnic2018apj} data and their different combinations. 

The paper is organised as follows. In section \ref{sec:bckgrnd} we briefly discuss the
background equations for the interacting dark energy model. The corresponding perturbation equations, the evolution of the density contrast, and the effects on the cosmic microwave background (CMB) temperature fluctuation and matter power spectrum have been provided in section \ref{sec:pert}. In section \ref{sec:obs-data}, the results obtained from constraining the DE model against different observational datasets performing the Markov Chain Monte Carlo (MCMC) analysis are discussed. Finally, conclusions are presented in section \ref{sec:sum}. 
\section{Dark matter-dark energy interaction} \label{sec:bckgrnd}
It is assumed that the Universe is described by a spatially flat, homogeneous and isotropic Friedmann-Lema\^itre-Robertson-Walker (FLRW) metric,
\begin{equation}\label{eq:metric}
ds^2= a^2(\tau)\paren*{- d \tau ^2+\delta_{ij} d x^i dx^j},
\end{equation}
where $a(\tau)$ is the scale factor in conformal time $\tau$ and the relation between conformal time ($\tau$) and cosmic time ($t$) is given as $a^2 d\tau^2 = dt^2$.  Using the metric (Eqn.\ (\ref{eq:metric})), the Friedmann equations are written as
\begin{eqnarray}
3 \cH^2 &=& a^2 \kappa \sum_{A}\rA \label{eq:fd1},\\ 
\cH^2+ 2 \cH^\prime &=& a^2 \kappa \sum_{A} \pA , \label{eq:fd2}
\end{eqnarray}
where $\kappa=8 \pi G_N$ ($G_N$ being the Newtonian Gravitational constant), $\cH\paren*{\tau}= \frac{a^\prime}{a}$ is the conformal Hubble parameter and $\rA$ and $\pA$ are respectively the energy density and pressure of the different components of the Universe. A prime indicates differentiation with respect to the conformal time $\tau$. It is assumed that the Universe is filled with photons ($\gamma$), neutrinos ($\nu$), baryons ($b$), cold dark matter ($c$) and dark energy ($de$). Among the different components, only cold dark matter and dark energy interact with each other and contribute to the energy budget together. The coupled conservation equations are
\begin{eqnarray}
\rdc^\prime+ 3 \cH \rdc &=& -aQ\,,\label{eq:con1}\\
\rde^\prime+ 3 \cH \paren*{1+\wde} \rde&=& aQ. \label{eq:con2}
\end{eqnarray}
where $Q$ gives the rate of energy transfer between the two fluids, $\wde = \pde/\rde$ is equation of state (EoS) parameter of dark energy and pressure, $p_{c} = 0$ for cold dark matter. The other three independent components --- photons ($\gamma$), neutrinos ($\nu$) and baryons ($b$) have their conservation equations as
\begin{equation}
\rho^\prime_{A}+ 3 \cH \paren*{1+w_{A}} \rA = 0\, \label{eq:con3},
\end{equation}
where $w_{A} = \pA/\rA$ is the EoS parameter of the $A$-th fluid and $A$ being $\gamma, \nu$ and  $b$. For photons and neutrinos, the EoS parameter is $w_{\gamma} = w_{\nu} = 1/3$, for baryons, the EoS parameter is $w_{b} = 0$. 

It is clear from Eqns.\ (\ref{eq:con1}) and (\ref{eq:con2}), if $Q<0$, energy flows from dark energy to dark matter (DE $\rightarrow$ DM) while if $Q>0$, energy flows from dark matter to dark energy (DM $\rightarrow$ DE). In literature~\cite{bohmer2008prd, valiviita2008jcap, gavela2009jcap, clemson2012prd, zhang2012jcap, acosta2014prd, yang2014prd1, das2015ass, divalentino2017prd1, divalentino2017prd2, yang2017prd1, yang2017prd2, pan2018mnras, yang2018prd, yang2018mnras, yang2018prd, yang2018jcap, vagnozzi2020mnras, sinha2020epjp,sinha2021prd}, the most popular forms of the phenomenological interaction term are $Q \propto \rdc$ or $Q \propto \rde$ or $Q$ proportional to any combination of them. Since there is no particular theoretical compulsion for the choice of the source term $Q$, we have considered the covariant form $Q^{\mu}$ such as
\begin{equation} \label{eq:inter}
Q^{\mu} =   \frac{\cH \rde \, u^{\mu}_{c} \, \beta}{a},
\end{equation}
with $\beta$ being the coupling parameter and $u^{\mu}_{c}$ being the cold dark matter 4-velocity. The coupling parameter determines the direction as well as amount of energy flow; $\beta = 0 $ indicates that the dark sector conserves independently. Thus $\beta > 0$ or equivalently $Q > 0$ will correspond to an evolution dynamics in which the dark matter component will redshift faster than $a^3$ and for $\beta < 0$ or $Q < 0$, the dark matter component will red shift at a rate slower than $a^3$ and hence should have its effect on the structure formation mechanism. Here we have considered that energy is transferred from DM to DE ($\beta>0$).

\begin{figure}[ht]
\begin{center}
\includegraphics[width=0.85\columnwidth]{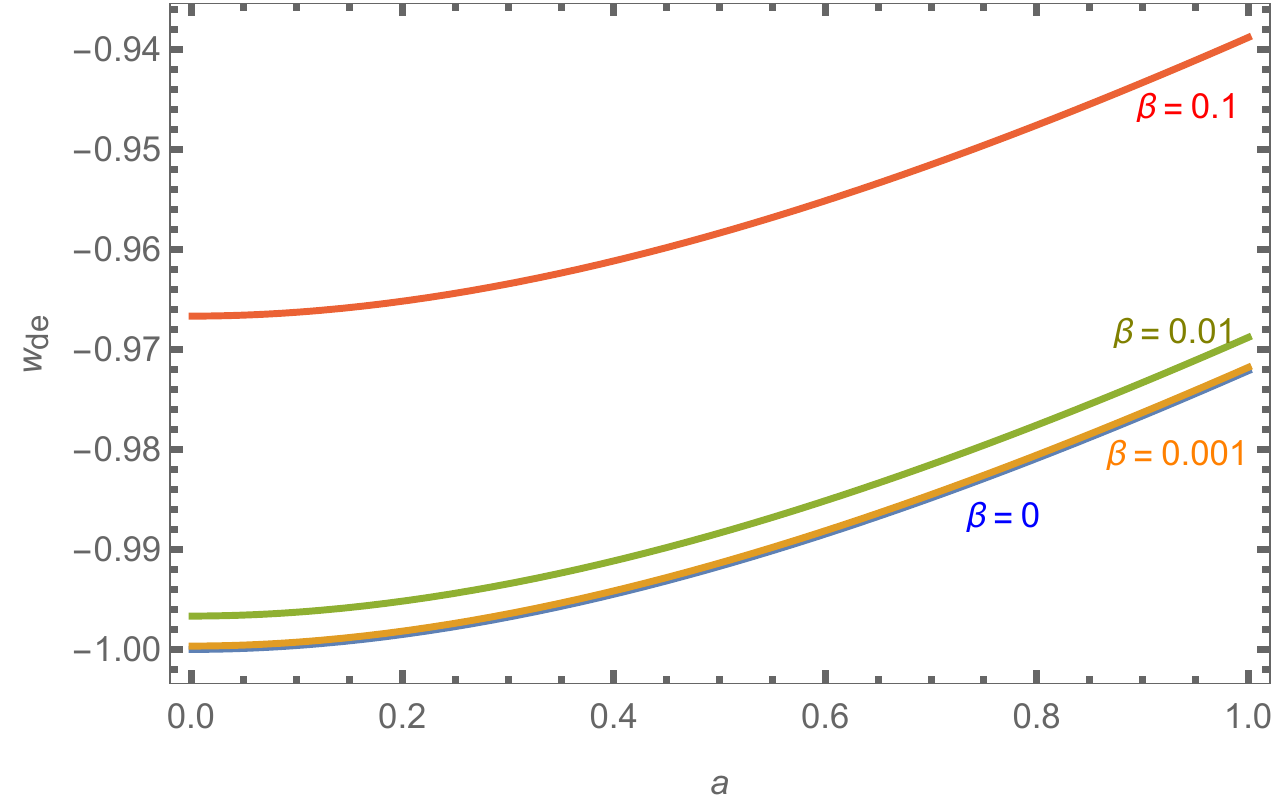}
\caption{The evolution of the dark-energy equation-of-state 
parameter $w_{de}(a)$, for different values of $\beta$. We have chosen $\gamma = 4.93$ and $\lambda = 2.94$ which are the best fit values obtained in \cite{Das2018RAA}.} 
\label{figwdevsa}
\end{center}
\end{figure}

For this work, instead of parametrising the EoS parameter directly, we have the chosen the ansatz for $\rde$, following Das \etal~\cite{Das2018RAA} as,
\begin{equation}\label{eq:rde}
 \frac{1}{\rho_{de}}\frac{d\rho_{de}}{da}=-\frac{\lambda a}{(\gamma+a)^2},
 \end{equation}
where $\lambda$ and $\gamma$ are constants. Solution of the differential equation (\ref{eq:rde}) gives the expression for $\rho_{de}$ as \cite{Das2018RAA}
\begin{equation}\label{rhode}
    \rho_{de}(a)=\frac{\rho_{de 0}(1+\gamma)^\lambda\exp{\left(\frac{\gamma\lambda}{1+\gamma}\right)}\exp{\left(-\frac{\gamma\lambda}{a+\gamma}\right)}}{(\gamma+a)^\lambda}
\end{equation}
For $\gamma = 0$, the above equation will provide a simple power law evolution of $\rde\paren*{a^{\lambda}}$, which has been considered in many cosmological analysis. It is well established that in order to facilitate the structure formation, an accelerating model of the Universe should have a preceding deceleration history. This particular parametrization ensures that $q(a)$ depicts an evolution history which allows the structure formation of the Universe to proceed unhindered. Using equation (\ref{eq:rde}), the parametrization for $\wde$ is obtained as
\begin{equation}\label{eq:wde}
\wde = -1 + \frac{\beta}{3} + \frac{a^{2}\lambda}{3\paren*{a+\gamma}^{2}}.
\end{equation}
Equation (\ref{eq:wde}) shows that for $\lambda\ll1$ and $\gamma\gg1$, $\wde\approx-1$ at the present epoch, even in presence of a small amount of energy flow from DM to DE. For the dark energy to produce the recent cosmic acceleration as well as to avoid the future ``big-rip'' singularity associated with phantom EoS parameter, the parameters ($\beta,\,\lambda,\,\gamma$) at $a=1$, must satisfy the condition,
\begin{equation}\label{eq:relation}
-\beta <\frac{\lambda}{\paren*{1+\gamma}^{2}} \leqslant 2-\beta.
\end{equation}
Figure \ref{figwdevsa} shows the variation of $w_{de}(a)$ for different values of the coupling parameter $\beta$. We have chosen here positive values of $\beta$ which correspond to energy flow from the dark matter to the dark energy sector and is observationally preferred direction of flow of energy \cite{yang2018prd, yang2018jcap, yang2018mnras, zhang2012jcap}. For Fig.\ (\ref{figwdevsa}), the model parameters $\gamma$ and $\lambda$ has been considered as $4.93$ and $2.94$ respectively which happens to be the best fit values obtained in \cite{Das2018RAA}. Now, as evident from the plot, in case of no interaction ($\beta=0$, denoted by blue curve in the graph) as well as when the strength of interaction is very less ( $\beta=0.001$, denoted by orange line), $w_{de}$ does not depart much from $\Lambda$CDM throughout the evolution as mentioned earlier. But as we go on increasing the strength of interaction, departure from $\Lambda$CDM becomes significant.  
Here the values of the coupling strength $\beta$ has been chosen arbitrarily. In most of the interacting dark energy models, the strength of interaction is considered to be very small such that there is no perturbational instability and also in order to be consistent with the observational results \cite{yang2014prd1, yang2017prd1,yang2017prd2, pan2018mnras, vagnozzi2020mnras}. In this work, for the particular phenomenological choice of interacting dark energy model given by equation (\ref{eq:rde}), we try to obtain constraints on various cosmological  parameters, particularly the coupling strength $\beta$. Also we are interested to know through perturbative analysis whether there is any instability in the growth of perturbations for this particular toy model.

For the background and perturbation analyses, we have considered three different combinations of the model parameters as given in table \ref{tab:parval}. In Case I, the values of ($\lambda,\,\gamma$) are chosen in close approximation with the best-fit values as obtained in~\cite{Das2018RAA} with a higher strength of interaction. In Case II, the strength of interaction is decreased considerably, consistent with the observational results obtained in~\cite{yang2014prd1, yang2017prd1, yang2017prd2, pan2018mnras, vagnozzi2020mnras, sinha2021prd} as well as the value of $\lambda$. In Case III, keeping $\beta$ same as the previous case, the value of $\gamma$ is decreased considerably but that of $\lambda$ is increased as well such that $\wde$ does not deviate a lot from $-1$ at the present epoch. 
\begin{table}[!htbp]
\begin{center}
\caption{\label{tab:parval}
Values of parameters used in this work.}
\begin{tabular}{cccc}
\hline \hline
Cases& \hspace{10ex}$\beta$& \hspace{10ex}$\lambda$& \hspace{10ex}$\gamma$\\
\hline
\rule[-1ex]{0pt}{2.5ex}Case I &\hspace{10ex}$0.10$&\hspace{10ex}$3.0$ &\hspace{10ex}$5.0$\\
\rule[-1ex]{0pt}{2.5ex}Case II &\hspace{10ex}$0.001$&\hspace{10ex}$0.3$ &\hspace{10ex}$5.0$\\
\rule[-1ex]{0pt}{2.5ex}Case III &\hspace{10ex}$0.001$&\hspace{10ex}$1.0$ &\hspace{10ex}$0.5$\\
\hline
\hline
\end{tabular}
\end{center}
\end{table}

\begin{figure*}[!htbp]
        \centering
            \subfloat{\includegraphics[width=.5\linewidth]{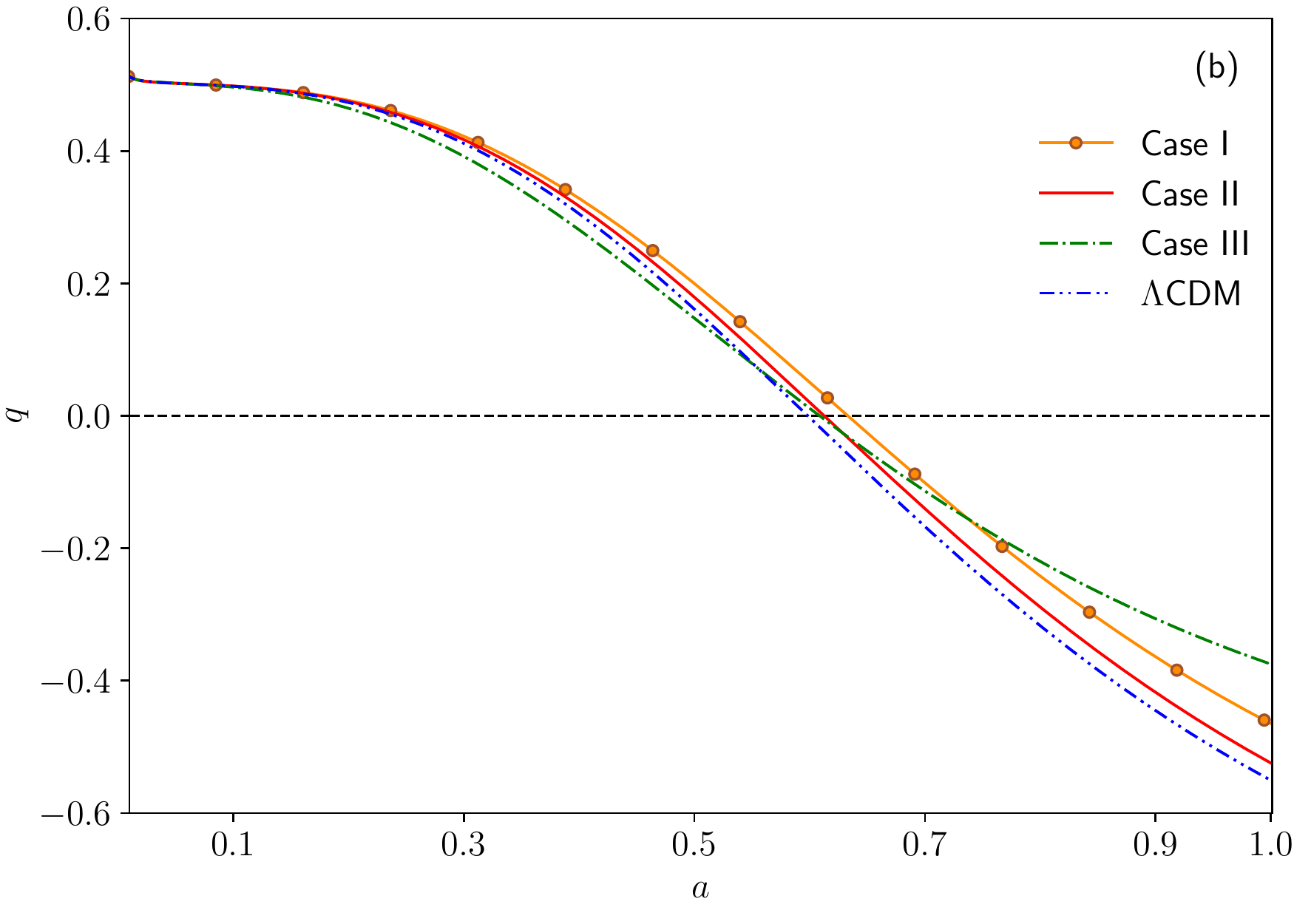}}\hfill
            \subfloat{\includegraphics[width=.5\linewidth]{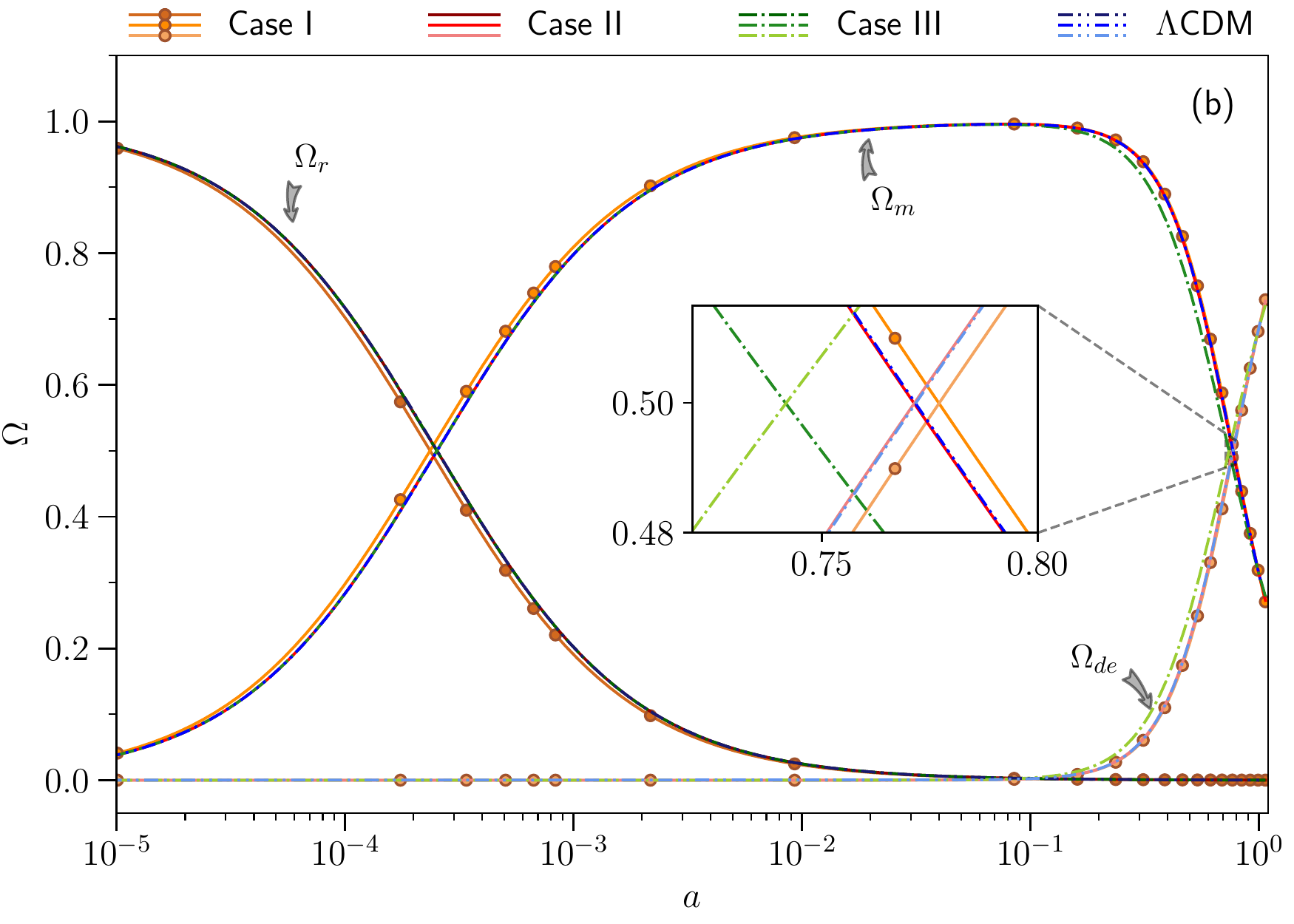}}\hfill
        \caption{Plot of (a) the dimensionless deceleration parameter $q$ and (b) density parameter $\Omega$ against scale factor $a$. The x-axis in Fig.\ (b) is in logarithmic scale. The solid line with solid circles represents Case I, solid line represents Case II and dashed-dot line represents Case III while the dashed-dot-dot line is for \lcdm. The inset shows the zoomed-in portion for the region $a= 0.76$ to $a = 0.80$.}\label{im:bck1}
\end{figure*}

To understand, how these parameter combinations affect the evolution of the Universe, we have shown the variation of the deceleration parameter, $q = -\paren*{\frac{a a^{\prime\prime}}{a^{\prime2}}-1}$  with the scale factor $a$ in Fig.\ (\ref{im:bck1}a). It is clear from Fig.\ (\ref{im:bck1}a), that the evolution of deceleration parameter is different than the \lcdm model even though $\wde$ is close to -1 at the present epoch. An interesting feature is that though the Universe starts to accelerate at the same epoch in Case II and Case III, the rate of acceleration is very different in the two cases. This confirms that, even for the same $\beta$ value, the combination of $\lambda,\,\gamma$ is crucial in determining the evolution history of the Universe. 

Figure (\ref{im:bck1}b) shows the evolution of the density parameters of radiation ($\Omega_{r}$), dark matter together with baryons ($\Omega_{m}$) and dark energy ($\Omega_{de}$) with scale factor $a$, in logarithmic scale. The density parameter of matter (baryonic matter and cold dark matter (DM), denoted as `$m\paren*{=b+c}$') is defined as $\Omega_{m}=\frac{\rdm}{3\,H^2 /\kappa}$ and that of dark energy (DE) is defined as $\Omega_{de}=\frac{\rde}{3\,H^2 /\kappa}$. Similarly, energy density parameter for radiation (denoted as `$r \paren*{=\gamma+\nu}$') is $\Omega_{r}=\frac{\rho_{r}}{3\,H^2 /\kappa}$. Here $H$ is the Hubble parameter defined with respect to the cosmic time $t$. The parameter values used in this section and in section \ref{sec:pert} are taken from the latest 2018 data release of the \Planck collaboration~\cite{planck2018cp}.
\section{Evolution of perturbations}\label{sec:pert}
The perturbed FLRW metric in a general gauge in conformal time is written as~\cite{kodama1984ptps, mukhanov1992pr, ma1995apj} 
\begin{equation} \label{eq:metric2}
\begin{split}
ds^2=a^2\paren*{\tau} & \left\{ -\paren*{1+2\phi}d\tau^2+2\,\partial_iB \,d\tau \,dx^i +\right. \\
& \left.\left[\paren*{1-2\psi}\delta_{ij}+2\partial_i\partial_jE\right]dx^idx^j \right\},
\end{split}
\end{equation}
where $\phi, \psi, B, E$ are gauge-dependant scalar functions of time and space. In presence of interaction, the fluids do not conserve independently and the covariant form of the energy-momentum conservation equation takes the form
\begin{equation} \label{eq:condition}
T^{\,\mu \nu}_{\paren*{A}; \nu} =Q^{\,\mu}_{\paren*{A}} \,, \hspace{0.3cm} \mbox{where}  \hspace{0.2cm} \sum_{A}Q^{\mu}_{\paren*{A}} =0 ~.
\end{equation}
Here $Q^{\mu}_{\paren*{A}}$ is the covariant form of the energy-momentum transfer function among the fluids~\cite{valiviita2008jcap, majerotto2010mnras, clemson2012prd}. Defining the $4$-velocity of fluid `$A$' as
\begin{equation}\label{eq:v4}
u^{\,\mu}_{A} = a^{-1}\paren*{1-\phi, v^{i}_{A}},
\end{equation}
with $v_{A}$ being the peculiar velocity of fluid `$A$', the covariant form of the source term (Eqn. \ref{eq:inter}), conveniently takes the form
\begin{equation} \label{eq:q1}
Q= \frac{\cH \rde\, \beta}{a}.
\end{equation}
Accounting for the pressure perturbation in presence of interaction~\cite{wands2000prd, malik2003prd, malik2005jcap, valiviita2008jcap, malik2009pr}, the perturbation conservation equations in Fourier space for dark matter and dark energy using synchronous gauge~\cite{ma1995apj} ($\phi=B=0$, $\psi=\eta$ and $k^2\,E=-\msh/2-3\eta$) are respectively written as
\begin{widetext}
\begin{eqnarray}
\ddc^\prime+ k v_{c} +\frac{\msh^\prime}{2} &=& \cH \beta \frac{\rde}{\rdc}\paren*{\ddc-\dde},  \label{eq:e2dm}\\
v_{c}^\prime+\cH v_{c}&=& 0~, \label{eq:m2dm}
\end{eqnarray}
\begin{equation}
\begin{split}
\dde^\prime +3  \cH & \paren*{\cde-\wde}\dde+\paren*{1+\wde}\paren*{k \vde+\frac{\msh^\prime}{2}}\\
+3 \cH & \left[3 \cH \paren*{1+\wde}\paren*{\cde-\wde}\right]\frac{\vde}{k} +3\cH \wde^\prime\frac{\vde}{k} \\
=\, 3 \cH^{2} & \beta  \paren*{\cde-\wde}\frac{\vde}{k}, \label{eq:e2de}
\end{split}
\end{equation}
\begin{equation}
\vde^\prime+\cH\paren*{1-3\cde}\vde-\frac{k\,\dde\,\cde }{\paren*{1+\wde}}=\frac{\cH\, \beta}{\paren*{1+\wde}}\, \left[\vdc-\paren*{1+\cde}\vde\right]. \label{eq:m2de}
\end{equation}
\end{widetext}
Here $\ddc = \delta\rdc/\rdc$ and $\dde = \delta\rde/\rde$ are the density contrasts of the dark matter and dark energy respectively, $\cde=\frac{\delta\pde}{\delta\rde}$ is the square of effective sound speed in the rest frame of DE, $k$ is the wavenumber, $\eta$ and $\msh$ are synchronous gauge fields in the Fourier space. For a detailed derivation of the perturbation equations, one may refer~\cite{kodama1984ptps, mukhanov1992pr, ma1995apj, wands2000prd, malik2003prd, malik2005jcap, valiviita2008jcap, malik2009pr, sinha2021}.

The coupled differential equations (Eqns.\ (\ref{eq:e2dm})-(\ref{eq:m2de})) are solved along with the perturbation equations~\cite{kodama1984ptps, mukhanov1992pr, ma1995apj} of the radiation, neutrino and baryon with $k = 0.1\,h$ $\mpci$ and the adiabatic initial conditions using the publicly available Boltzmann code \camb\footnote{Available at: \href{https://camb.info}{https://camb.info}}~\cite{lewis1999bs} after suitably modifying it. 

The adiabatic initial conditions for $\ddc$, $\dde$ in presence of interaction are respectively
\begin{subequations}
\begin{eqnarray}
\delta_{ci} &=& \left[3 +\frac{\rde}{\rdc}\beta\right] \frac{\delta_{\gamma}}{3\paren*{1+w_{\gamma}}}, \label{eq:initial-m}\\
\delta_{dei} &=& \left[3\,\paren*{1+\wde} - \beta \right] \frac{\delta_{\gamma}}{3\paren*{1+w_{\gamma}}}, \label{eq:initial-de}
\end{eqnarray}
\end{subequations}
where, $w_{\gamma}$ is the EoS parameter and $\delta_{\gamma}$ is the density fluctuation of photons. As can be seen from Eqn.\ (\ref{eq:m2dm}), there is no momentum transfer in the DM frame, hence initial value for $v_{c}$ is set to zero ($v_{ci} =0$)~\cite{bean2008prd2, chongchitnan2009prd, xia2009prd, valiviita2008jcap}. The initial value for the dark energy velocity, $\vde$ is assumed to be same as the initial photon velocity, $v_{dei} = v_{\gamma\,i}$.
To avoid the instability in dark energy perturbations due to the the propagation speed of pressure perturbations, we have set $\cde = 1$~\cite{waynehu1998apj, bean2004prd, gordon2004prd, afshordi2005prd, valiviita2008jcap}.

\begin{figure*}[!htbp]
  \centering
\includegraphics[width=\textwidth]{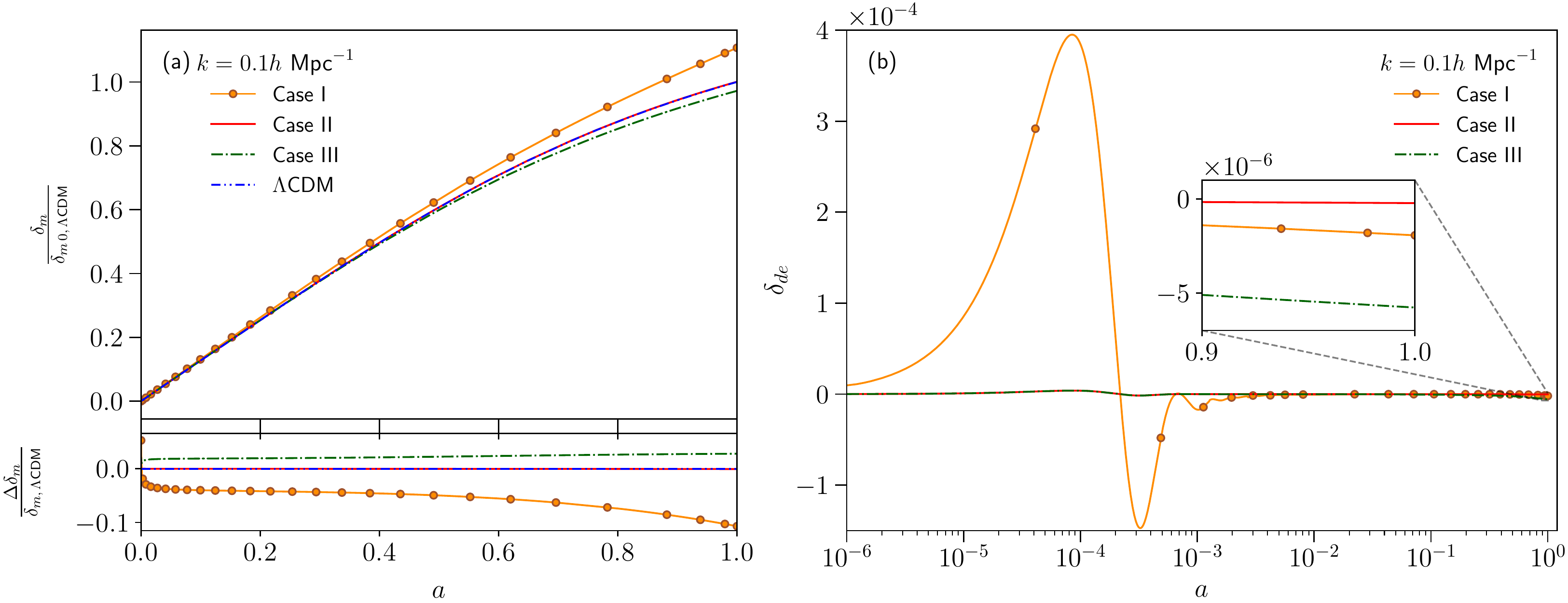}
\caption{(a) Plot of \textbf{Upper Panel} : the matter density contrast $\frac{\delta_{m}}{\delta_{m\,0,\,\scriptsize{\Lambda\text{CDM}}}}$ and \textbf{Lower Panel} : fractional growth rate is defined as $\frac{\Delta \delta_{m}}{\delta_{m,\,\scriptsize{\Lambda\text{CDM}}}} = \paren*{1- \frac{\delta_{m}}{\delta_{m,\,\scriptsize{\Lambda\text{CDM}}}}}$ relative to the \lcdm model against $a$. The origin on the x-axis represents $10^{-6}$. (b) Plot of the dark energy density fluctuation, $\dde$ against $a$ in logarithmic scale for $k = 0.1\,h$ $\mpci$. The inset shows the zoomed-in portion from $a = 0.9$ to $a = 1.0$. The solid line with solid circles represents Case I, solid line represents Case II and dashed-dot line represents Case III while the dashed-dot-dot line is for \lcdm.}\label{im:delta1} 
\end{figure*}

Figure (\ref{im:delta1}a) shows the variation of the matter density contrast, $\ddm = \delta \rdm/\rdm$ which includes both the cold dark matter ($c$) and the baryonic matter ($b$) against $a$ for different test values of the model parameters given in table \ref{tab:parval}. The matter density contrast for the \lcdm model is also shown. For a better comparison with the \lcdm model, $\ddm$ is scaled by $\delta_{m0} = \ddm\paren*{a=1}$ of \lcdm\footnote{The origin on the x-axis is actually $10^{-6}$}. In Case I, the growth is very close to that of \lcdm model at early times; when the effect of interaction comes into play at late time, the growth rate increases and $\ddm$ reaches a higher value compared to the \lcdm counterpart. In Case II, the growth of density fluctuation $\ddm$ is exactly like the \lcdm model. In Case III, the Universe decelerates faster resulting in lesser growth of $\ddm$ compared to the \lcdm model. To understand better how the different model parameters affect the growth of perturbations, fractional matter density contrast, $\frac{\Delta \delta_{m}}{\delta_{m,\,\scriptsize{\Lambda\text{CDM}}}} = \paren*{1- \frac{\delta_{m}}{\delta_{m,\,\scriptsize{\Lambda\text{CDM}}}}}$ are shown in the lower panel of Fig.\ (\ref{im:delta1}a). Figure (\ref{im:delta1}b) shows the variation of the dark energy density contrast $\dde$ for the different cases of the interacting model. At early time, $\dde$ oscillates and then decays to very small values. The amplitude of oscillation is higher for higher $\beta$ value. 


\subsection{Effect on CMB temperature and matter power spectrum}\label{sec:result}
It is useful to understand how the different model parameters affect the CMB temperature spectrum, matter power spectrum. We have computed the CMB temperature power spectrum ($C_{\ell}^{TT}$) and the matter power spectrum ($P\left(k,a\right)$) numerically using our modified version of \camb. For a detailed analysis on power spectra one may refer to~\cite{hu1995apj,seljak1996apj,dodelson2003}.
\begin{figure*}[!htbp]
\centering
\includegraphics[width=\textwidth]{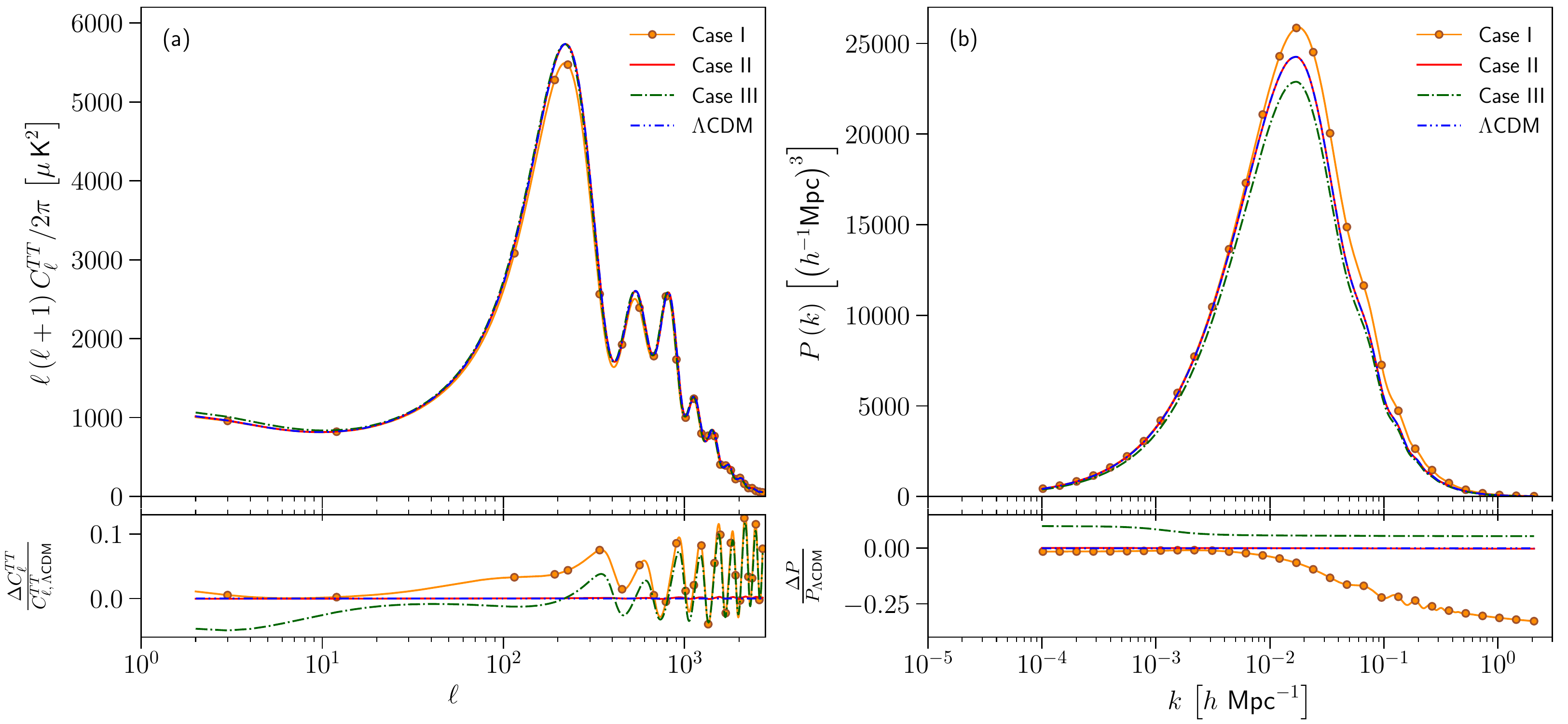}
\caption{\textbf{Upper Panel} : (a) Plot of CMB temperature power spectrum in units of $\mu \mbox{K}^{2}$ with the multipole index $\ell$ in logarithmic scale. (b) Plot of matter power spectrum $P\paren*{k}$ in units of $\paren*{h^{-1}\mpc}^{3}$ with wavenumber $k$ in units of $h\,\mpci$. \textbf{Lower Panel} : Plot of fractional change in the temperature spectrum, $\frac{\Delta C_{\ell}^{TT}}{C_{\ell,\, \scriptsize{\Lambda\text{CDM}}}^{TT}} = \paren*{1- \frac{C_{\ell}^{TT}}{C_{\ell,\, \scriptsize{\Lambda\text{CDM}}}^{TT}}}$ and the fractional change in matter power spectrum, $\frac{\Delta P}{P_{\scriptsize{\Lambda\text{CDM}}}} = \paren*{1- \frac{P}{P_{\scriptsize{\Lambda\text{CDM}}}}}$. For both the panels, solid line with solid circles represents Case I, solid line represents Case II and dashed-dot line represents Case III while the dashed-dot-dot line is for the \lcdm at $a=1$.}\label{im:cl} 
\end{figure*}

Figure (\ref{im:cl}a) shows the temperature power spectrum for the three test cases and \lcdm at the present epoch, $a=1$. The lower panel of Fig.\ (\ref{im:cl}a), shows the fractional change ($=\Delta C_{\ell}^{TT}/C_{\ell,\, \scriptsize{\Lambda\text{CDM}}}^{TT}$) in $C_{\ell}^{TT}$ to emphasize the difference between the various cases and the \lcdm model. It is clear that Case II behaves exactly like the \lcdm model. In Case I, deviation from the \lcdm model is higher around the first peak, where as in Case III, the effect comes due to the integrated Sachs-Wolfe (ISW) effect. As seen from Eqn. (\ref{eq:wde}), for negligible interaction, on decreasing $\gamma$, the EoS parameter $\wde$ deviates more from $-1$ and significantly changes the higher multipole behaviour.

Figure (\ref{im:cl}b) shows the matter power spectrum for the three test cases and \lcdm at the present epoch, $a=1$. In Case I, the matter-dark energy equality and hence the epoch of acceleration is more towards the recent past, giving the matter perturbation enough time to cluster more. This is manifested as higher value of $P\paren*{k}$ for smaller modes compared to the \lcdm model. Case II behaves exactly like the \lcdm model.  In Case III, the matter-dark energy equality is more towards the past resulting in a lesser clustering of matter and hence lower value of $P\paren*{k}$ for smaller modes. Moreover, in Case III, the expansion rate of the Universe is very different from the other cases resulting in a lesser clustering even for larger scales. These features are clear from the lower panel of Fig.\ (\ref{im:cl}b), which shows the  fractional change in matter power spectrum, $\Delta P/P_{\scriptsize{\Lambda\text{CDM}}}$ of the test cases relative to the \lcdm model. 
\section{Observational Constraints}\label{sec:obs-data}
We obtain the observational constraints on the interacting dark energy model in this section. For that, the publicly available observational datasets used are the following:
\begin{description}
\item[CMB]
The latest 2018 data release of the \Planck collaboration\footnote{Available at: \href{http://pla.esac.esa.int/pla/\#home}{ https://pla.esac.esa.int}}~\cite{planck2019cmb,planck2018cp} for the cosmic microwave background (CMB) anisotropies data. The likelihoods considered are combined temperature (TT), polarization (TE) and temperature-polarization (EE) likelihood along with the CMB lensing likelihood (\plancklensing). The likelihoods together are represented as \Planck in results given in Sec.\ \ref{sec:obs-results}.

\item[BAO]
Datasets from the three surveys, the 6dF Galaxy Survey (6dFGS) measurements~\cite{beutler2011mnras} at redshift $z = 0.106$, the Main Galaxy Sample of Data Release $7$ of the Sloan Digital Sky Survey (SDSS-MGS)~\cite{ross2015mnras} at redshift $z = 0.15$ and the latest Data Release $12$ (DR12) of the Baryon Oscillation Spectroscopic Survey (BOSS) of the Sloan Digital Sky Survey (SDSS) III at redshifts $z = 0.38$, $0.51$ and $0.61$~\cite{alam2017sdss3}, are considered for the baryon acoustic oscillations (BAO) data. 

\item[Pantheon] 
For the luminosity distance measurements of the Type Ia supernovae (SNe Ia) measurements the compilation of 276 supernovae discovered by the Pan-STARRS1 Medium Deep Survey at $0.03 < z < 0.65$ and various low redshift and Hubble Space Telescope (HST) samples to give a total of 1048 supernovae data at $0.01 < z < 2.3$, called the `Pantheon' catalogue is used~\cite{scolnic2018apj}.
\end{description}

The datasets are used to constrain the six-dimensional parameter space of the \lcdm model and the three model parameters. The nine-dimensional parameter space is written as
\begin{equation}\label{eq:parameter}
\mathcal{P} \equiv \lbrace \Omega_{b} h^2, \Omega_{c} h^2, 100\theta_{MC}, \tau, \beta, \lambda,\gamma, \ln\paren*{10^{10} A_s}, n_{s}\rbrace,
\end{equation}
where $\Omega_b h^2$ is the baryon density, $\Omega_c h^2$ is the cold dark matter density, $\theta_{MC}$ is the angular acoustic scale, $\tau$ is the optical depth, $\beta$, $\lambda$ and $\gamma$ are the free model parameters, $A_{s}$ is the scalar primordial power spectrum amplitude and $n_{s}$ is the scalar spectral index. The posterior distribution of the parameters is sampled using the Markov Chain Monte Carlo (MCMC) simulator through a suitably modified version of the publicly available code \cosmomc\footnote{Available at: \href{https://cosmologist.info/cosmomc/}{https://cosmologist.info/cosmomc/}}~\cite{lewis2013hha,lewis2002ah}. For the statistical analysis, flat priors ranges are considered for all the parameters, given in Table \ref{tab:prior}. The statistical convergence of the MCMC chains is set to satisfy the Gelman and Rubin criterion~\cite{gelman1992}, $R-1 \lesssim 0.01$.

\begin{table}[!h]
\begin{center}
\caption{Prior ranges of nine independent parameters considered in the \cosmomc analysis.}\label{tab:prior}
\begin{tabular}{cc}
\hline \hline
\rule[-1ex]{0pt}{2.5ex}Parameter &  \hspace{24ex} Prior\\
\hline
\rule[-1ex]{0pt}{2.5ex}$\Omega_{b} h^2$&  \hspace{24ex} $\left[0.005, 0.1\right]$ \\
\rule[-1ex]{0pt}{2.5ex}$\Omega_{c} h^2$&  \hspace{24ex} $\left[0.001, 0.99\right]$\\
\rule[-1ex]{0pt}{2.5ex}$100\theta_{MC}$&  \hspace{24ex} $\left[0.5, 10\right]$\\
\rule[-1ex]{0pt}{2.5ex}$\tau$&  \hspace{24ex} $\left[0.01, 0.8\right]$\\
\rule[-1ex]{0pt}{2.5ex}$\beta$&  \hspace{24ex} $\left[-0.001, 1.0\right]$\\
\rule[-1ex]{0pt}{2.5ex}$\lambda$&  \hspace{24ex} $\left[0.0, 40.0\right]$\\
\rule[-1ex]{0pt}{2.5ex}$\gamma$&  \hspace{24ex} $\left[0.0, 40.0\right]$\\
\rule[-1ex]{0pt}{2.5ex}$\ln\paren*{10^{10} A_s}$&  \hspace{24ex} $\left[1.61, 3.91\right]$\\
\rule[-1ex]{0pt}{2.5ex}$n_{s}$&  \hspace{24ex} $\left[0.8, 1.2\right]$\\
\hline \hline
\end{tabular}
\end{center}
\end{table}
\subsection{Observational results}\label{sec:obs-results}
%
\begin{center}
\begin{table*}[!htbp]
\centering
\caption{Observational constraints on the nine dependent model parameters with three derived parameters separated by a horizontal line and the error bars correspond to $68\%$ confidence level, using different observational datasets.}
\label{tab:mean-1}
\begin{tabular} { l  c c c}
\hline\hline \noalign{\vskip 2pt}
 {\normalsize Parameter} &\hspace{7ex}  {\normalsize \Planck} &\hspace{7ex}  {\normalsize \Planck \dataplus BAO} &\hspace{7ex}  {\normalsize \thead{\Planck \\ \dataplus BAO \dataplus Pantheon}}\\
\hline \noalign{\vskip 2pt}
\rule[-1.5ex]{0pt}{2.7ex}{\boldmath$\Omega_b h^2   $} &\hspace{7ex} $0.022411\pm 0.000157      $ &\hspace{7ex} $0.022468\pm 0.000146      $ &\hspace{7ex} $0.022479\pm 0.000145      $\\
\rule[-1.5ex]{0pt}{2.7ex}{\boldmath$\Omega_c h^2   $} &\hspace{7ex} $0.11960\pm 0.00120        $ &\hspace{7ex} $0.118875\pm 0.000959      $ &\hspace{7ex} $0.118764\pm 0.000923      $\\
\rule[-1.5ex]{0pt}{2.7ex}{\boldmath$100\theta_{MC} $} &\hspace{7ex} $1.040795\pm 0.000315      $ &\hspace{7ex} $1.040879\pm 0.000298      $ &\hspace{7ex} $1.040886\pm 0.000297      $\\
\rule[-1.5ex]{0pt}{2.7ex}{\boldmath$\tau           $} &\hspace{7ex} $0.05338\pm 0.00724        $ &\hspace{7ex} $0.05514\pm 0.00740        $ &\hspace{7ex} $0.05546^{+0.00675}_{-0.00756}$\\
\rule[-1.5ex]{0pt}{2.7ex}{\boldmath$\beta          $} &\hspace{7ex} $0.00757^{+0.00276}_{-0.00657}$ &\hspace{7ex} $0.00781^{+0.00288}_{-0.00672}$ &\hspace{7ex} $0.00790^{+0.00303}_{-0.00661}$\\
\rule[-1.5ex]{0pt}{2.7ex}{\boldmath$\lambda        $} &\hspace{7ex} --- &\hspace{7ex} --- &\hspace{7ex} $< 24.6                    $\\
\rule[-1.5ex]{0pt}{2.7ex}{\boldmath$\gamma         $} &\hspace{7ex} $> 15.9                    $ &\hspace{7ex} $> 17.3                    $ &\hspace{7ex} $> 20.7                    $\\
\rule[-1.5ex]{0pt}{2.7ex}{\boldmath${\rm{ln}}(10^{10} A_s)$} &\hspace{7ex} $3.0454^{+0.0131}_{-0.0146}$ &\hspace{7ex} $3.0482\pm 0.0143          $ &\hspace{7ex} $3.0485\pm 0.0145          $\\
\rule[-1.5ex]{0pt}{2.7ex}{\boldmath$n_s            $} &\hspace{7ex} $0.96378\pm 0.00428        $ &\hspace{7ex} $0.96540\pm 0.00393        $ &\hspace{7ex} $0.96569\pm 0.00393        $\\
\hline \noalign{\vskip 2pt}
\rule[-1.5ex]{0pt}{2.7ex}$H_0                       $ &\hspace{7ex} $66.94^{+1.02}_{-0.397}    $ &\hspace{7ex} $67.473^{+0.607}_{-0.412}  $ &\hspace{7ex} $67.660^{+0.464}_{-0.421}  $\\
\rule[-1.5ex]{0pt}{2.7ex}$\Omega_m                  $ &\hspace{7ex} $0.31881^{+0.00590}_{-0.0122}$ &\hspace{7ex} $0.31197^{+0.00559}_{-0.00710}$ &\hspace{7ex} $0.31000\pm 0.00571        $\\
\rule[-1.5ex]{0pt}{2.7ex}$\sigma_8                  $ &\hspace{7ex} $0.8103^{+0.0102}_{-0.00530}$ &\hspace{7ex} $0.81167^{+0.00784}_{-0.00670}$ &\hspace{7ex} $0.81280\pm 0.00688        $\\
\hline\hline
\end{tabular}
\end{table*}
\end{center}
\begin{figure*}[!htbp]
        \centering
         \includegraphics[width=0.75\linewidth]{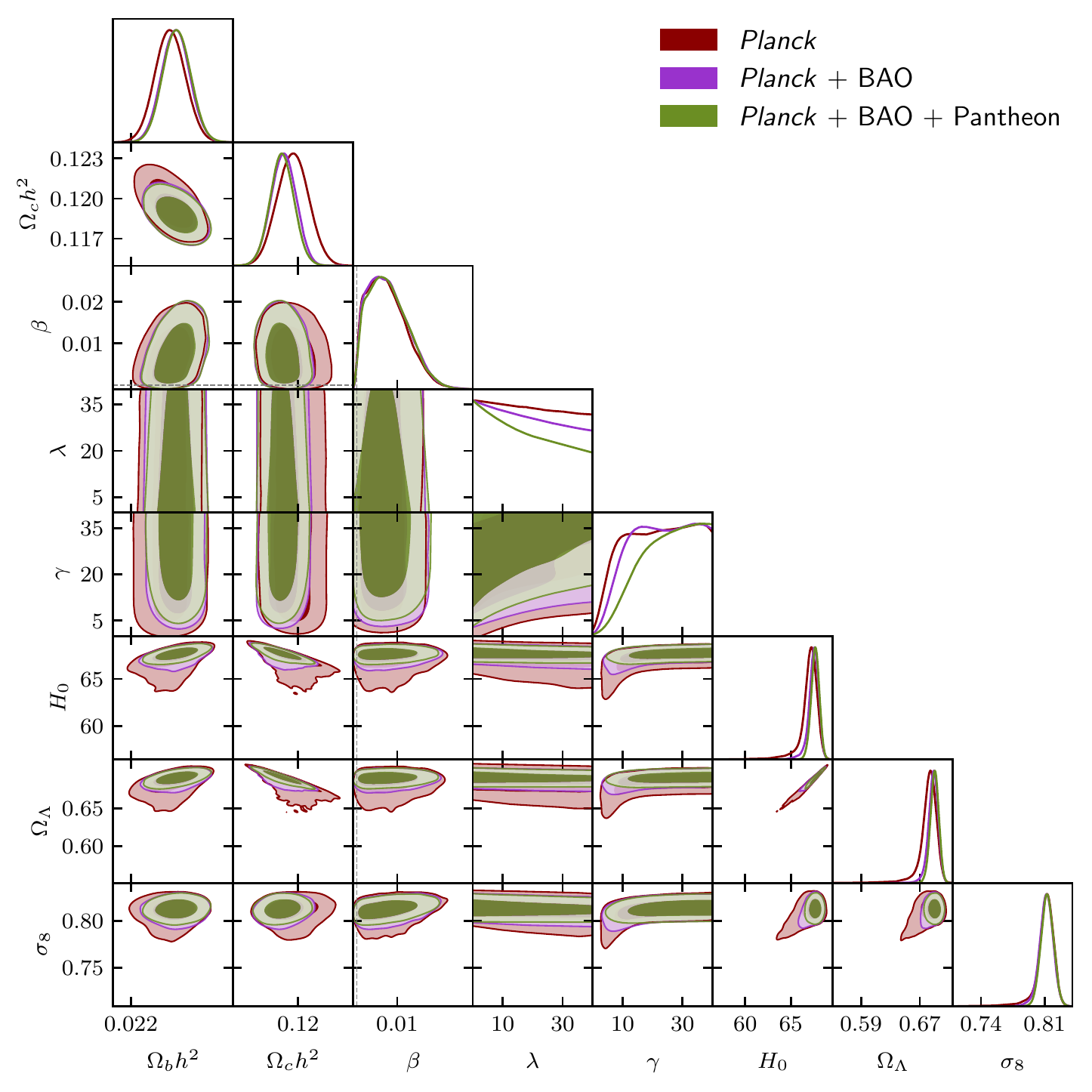}
         \caption{Plot of 1-dimensional marginalised posterior distributions and 2-dimensional marginalised constraint contours on the parameters containing $68\%$ and $95\%$ probability.}\label{im:tri4}
\end{figure*}

The marginalised values with errors at $1\sigma$ ($68\%$ confidence level) of the nine free parameters and three derived parameters, $H_{0}$, $\Omega_{m}$ and $\se$, are listed in Table \ref{tab:mean-1}. When only the \Planck data is considered, the central value of the coupling parameter, $\beta(=0.00757^{+0.00276+0.00898}_{-0.00657-0.00769})$, is non-zero within the $1\sigma$ region. The non-interacting case ($\beta=0$) lies in the $2\sigma$ region. The model parameter $\lambda$ has no boundary in the used prior range. The parameter $\gamma (> 15.9>5.30)$ has only the lower boundary both in $1\sigma$ and $2\sigma$ regions. The other parameters agree well with the values obtained from the \Planck estimation~\cite{planck2018cp}. Tensions in the central values of $H_{0}$ and $\se$ with direct measurements persist in this interacting DE model.

Addition of the BAO to the \Planck data, changes the central value of $\beta$ negligibly to $0.00781^{+0.00288+0.00907}_{-0.00672-0.00785}$ with zero inside the $2\sigma$ region. The \Planck \dataplus BAO combination provides no constrain on $\lambda$ and an increased lower limit on $\gamma(>17.3>7.29)$. The central value of the Hubble parameter increases very little to $67.473^{+0.607}_{-0.412}$. The error bars on $H_{0}$ decrease considerably on addition of the BAO data. 

Addition of the Pantheon data to the \Planck \dataplus BAO combination, affects the central value of $\beta (=0.00790^{+0.00303+0.00898}_{-0.00661-0.00799})$ very less. For the combined datasets also, $\beta=0$ lies in the $2\sigma$ region. The combined dataset can provide an upper limit on the parameter $\lambda(<24.6)$. The lower limit on $\gamma(>20.7>9.76)$ further increases. The other parameters do not change significantly on addition of different datasets.

In Fig.\ \ref{im:tri4}, the correlations between the parameters ($\Omega_b h^2$, $\Omega_c h^2$,  $\beta$, $\lambda$, $\gamma$) and the derived parameters ($H_{0}$, $\Omega_{\Lambda}$, $\se$) and their marginalised contours are shown. The contours contain $1\sigma$ region ($68\%$ confidence level) and $2\sigma$ region ($95\%$ confidence level). The coupling parameter $\beta$ is very slightly positively correlated to $\se$ but remains uncorrelated to other parameters. Figure \ref{im:tri4} highlights that the parameters ($\lambda,\,\gamma$) are uncorrelated with the other parameters. The parameter $\se$ is positively correlated with $H_{0}$ and $\Omega_{\Lambda}$ and remains uncorrelated to $\Omega_b h^2$ and $\Omega_c h^2$. Strong positive correlation of $\Omega_{\Lambda}$ with $H_{0}$ is clear from Fig.\ \ref{im:tri4}. $\Omega_b h^2$ remains positively correlated to $H_{0}$ and $\Omega_{\Lambda}$, whereas $\Omega_c h^2$ remains negatively correlated to $H_{0}$ and $\Omega_{\Lambda}$. 
%
\section{Summary and Discussion} \label{sec:sum}
In this paper, we have considered a modified form of the DE model described in \cite{Das2018RAA}. The modification has been introduced because, in this particular DE model, the dark sectors were not allowed to evolve independently; rather, the two dark sectors were considered to interact through a dynamical coupling term. The nonzero coupling term will affect the evolution of the dark sectors and should have its imprint on the growth of perturbations. We have studied the perturbative effect of this particular interacting DE model. It has been found that there was no significant effect on the matter density fluctuation for a lower rate of interaction. However, with the increase in the strength of interaction of the coupling term, dark energy density fluctuations exhibited visible imprints in the early epochs of evolution. 

We have worked out a detailed perturbation analysis in the synchronous gauge for different parameter values. We have also computed the CMB temperature spectrum and matter power spectrum. From the perturbation analysis, we have noted that, through appropriate tuning of the model parameters, we can obtain perturbation evolution almost identical to the \lcdm model, even for different background dynamics. 

We have tested the interacting model against the recent observational datasets like CMB, BAO and Pantheon with the standard six parameters of \lcdm model and the three model parameters, $\beta$, $\lambda$ and $\gamma$. We have obtained the central value of the coupling parameter, $\beta$ to be positive, indicating an energy flow from dark matter to dark energy. For all the datasets, $\beta=0$ lies outside the $1\sigma$ error region. We have considered a large prior range for $\lambda$ and $\gamma$ so that The priors of $\gamma$ and $\lambda$ are set such that the condition (Eqn.\ (\ref{eq:relation})) is satisfied and $\wde$ remains close to the \lcdm value. The parameters $\lambda$ and $\gamma$ are not constrained properly by the datasets used. Thus, we conclude from the perturbation analysis and the observational constraints that the model resolves the coincidence problem and produces an evolution dynamics close to the \lcdm model for any small value of $\lambda$ as long as $\gamma$ is large enough. As per the available data, higher values of interaction rate are not preferred much as this leads to additional features in the power spectrum, but the future surveys may result in a different perspective.
\section*{Acknowledgement}

The authors would like to acknowledge the use of ``Dirac Supercomputing Facility''\footnote{Details available at: \href{https://www.iiserkol.ac.in/~dirac/index.html}{https://www.iiserkol.ac.in/~dirac/index.html}} of IISER Kolkata. SD would like to acknowledge  IUCAA,  Pune  for  providing  support  through  the associateship programme.  

\numberwithin{equation}{section}
\numberwithin{table}{section}
%

%

%

\end{document}